\documentclass[aip,jcp,reprint,amsmath,amssymb,floatfix,citeautoscript]{revtex4-1}
\usepackage{amsmath}
\usepackage{graphicx}
\usepackage{amssymb}
\usepackage{amsthm}
\usepackage{bm}
\usepackage{subfig}
\usepackage{physics}

\usepackage{ragged2e}
\usepackage[labelsep=period,format=plain,justification=justified,font=small]{caption}
\DeclareCaptionJustification{myjust}{\justifying}
\usepackage{txfonts}

\usepackage{color}
\usepackage[usenames,dvipsnames]{xcolor}
\definecolor{myblue}{rgb}{0,0,1}
\usepackage[breaklinks=true,colorlinks=true,linkcolor=myblue,urlcolor=myblue,citecolor=myblue]{hyperref}

\begin{document}

\title{Linear and nonlinear spectroscopy from quantum master equations}

\author{Jonathan H. Fetherolf}
\author{Timothy C. Berkelbach}
\email{berkelbach@uchicago.edu}
\affiliation{Department of Chemistry and James Franck Institute,
University of Chicago, Chicago, Illinois 60637 USA}

\begin{abstract}
We investigate the accuracy of the second-order time-convolutionless (TCL2)
quantum master equation for the calculation of linear and nonlinear
spectroscopies of multichromophore systems.  We show that, even for systems
with non-adiabatic coupling, the TCL2 master equation predicts linear
absorption spectra that are accurate over an extremely broad range of
parameters and well beyond what would be expected based on the perturbative
nature of the approach; non-equilibrium population dynamics calculated with
TCL2 for identical parameters are significantly less accurate.  For third-order
(two-dimensional) spectroscopy, the importance of population dynamics and the
violation of the so-called quantum regression theorem degrade the accuracy of
TCL2 dynamics.  To correct these failures, we combine the TCL2 approach with a
classical ensemble sampling of slow microscopic bath degrees of freedom,
leading to an efficient hybrid quantum-classical scheme that displays excellent
accuracy over a wide range of parameters.  In the spectroscopic setting, the
success of such a hybrid scheme can be understood through its separate
treatment of homogeneous and inhomogeneous broadening.  Importantly, the
presented approach has the computational scaling of TCL2, with the modest
addition of an embarrassingly parallel prefactor associated with ensemble sampling.
The presented approach can be understood as a generalized inhomogeneous
cumulant expansion technique, capable of treating multilevel systems with
non-adiabatic dynamics.
\end{abstract}

 \maketitle

\section{Introduction}
\label{sec:intro}

Non-adiabatic energy transfer in molecular systems represents a
problem of broad interest in chemistry, physics, and materials science.  
In the condensed phase, these processes commonly occur with many comparable
energy scales, precluding simple perturbative treatments of the dynamics such
as Golden-rule-type rate theories.  This class of problems has motivated the
important development of accurate numerical techniques capable of evolving the
electronic reduced density matrix and offering insight into the population
dynamics of multi-level dissipative quantum
systems~\cite{Tanimura1989,Mak1990,Makarov1994,Beck2000,Thoss2001,Dunkel2008,Chen2017}.
However, with few exceptions, elements of the reduced density matrix are
basis-dependent and not directly observable.  Instead, the principal
experimental probes of energy transfer dynamics are ultrafast time-resolved
spectroscopies, such as pump-probe transient absorption and coherent
two-dimensional
spectroscopy~\cite{MukamelBook,Hybl1998,Mukamel2000,Jonas2003,Brixner2005,Engel2007}.
In this manuscript, we evaluate the accuracy of perturbative, but
non-Markovian, quantum master equations for the calculation of linear and
nonlinear spectroscopies.  In particular, we focus on systems characteristic of
protein-protected biological chromophores. This class of problems has been the
topic of intense study in the quantum dynamics
community~\cite{Jang2008,Plenio2008,Mohseni2008,Huo2010,Wu2010,Kolli2011,Tiwari2013,Tempelaar2014},
in part because the environmental fluctuations exhibit long correlation times
that challenge conventional theories~\cite{Ishizaki2009a,Ishizaki2009b}.

The Hamiltonian for multichromophore systems can be generically written as the
sum of an electronic (system) Hamiltonian, a nuclear (bath) Hamiltonian, and
the interaction between the two, $H=H_{\mathrm{s}}+H_{\mathrm{b}}+V$.  
In the present manuscript, we consider a Frenkel exciton model of coupled chromophores,
with the system Hamiltonian
\begin{equation}
    H_{\mathrm{s}} = \sum_m \left(\overline{\epsilon}+\epsilon_{m}\right) B^{\dagger}_{m} B_{m}
        + \sum_{mn}J_{mn} B^{\dagger}_{m}B_{n},
\label{eq:hamsys}
\end{equation}
where $\overline{\epsilon}$ is a mean excitation energy for the excited-state manifold
(equal to 10,000--30,000~cm$^{-1}$ for visible-light absorbing chromophores),
$\epsilon_{m}$ is the deviation from this mean excitation energy for site $m$, and $J_{mn}$
is the electronic coupling between sites $m$ and $n$. The operators $B^{\dagger}_{m}$ and
$B_{m}$ create and annihilate localized excitations on site $m$ and satisfy the 
commutation relation $[B_{m},B^{\dagger}_{n}]=\delta_{mn}(1-2B^{\dagger}_{m}B_{m})$. 
The bath Hamiltonian is that of the nuclear degrees of freedom in the
electronic ground-state.  The system-bath interaction can be generically decomposed into
the form $V = \sum_a E_a F_a$ where $E_a$ are bath operators and $F_a$ are
system operators. For simplicity, we consider the case
\begin{equation}
V = \sum_m E_m B_m^\dagger B_m,
\label{eq:hamsysbath}
\end{equation}
where $E_m$ is a collective bath operator whose fluctuations act to modulate the energy
gap of molecular site $m$.
An otherwise generic second-order perturbation theory in the system-bath interaction
requires only the equilibrium time correlation function of the nuclear degrees of freedom
in the electronic ground state,
\begin{equation}
\label{eq:gapCt}
C_m(t) = \mathrm{Tr}_{\mathrm{b}} \left\{ E_m(t) E_m(0) \rho_{\mathrm{b}}^{\mathrm{eq}} \right\}
\end{equation}
where $E_m(t) = \exp(iH_\mathrm{b}t/\hbar) E_m \exp(-iH_\mathrm{b}t/\hbar)$.
For simplicity, we assume nuclear degrees of freedom belonging to different
molecular sites are uncorrelated.
This approach is commonly pursued to account for dephasing dynamics via
atomistic simulations; in this approach, classical molecular dynamics are used to generate
trajectories for the evaluation of the energy gap autocorrelation function 
in Eq.~(\ref{eq:gapCt}), leading to the second-order cumulant approximation to the 
lineshape~\cite{Kubo1969,MukamelBook},
\begin{equation}
I_m(\omega) \propto \int_0^\infty dt e^{i\omega t} 
    \exp\left[-\int_0^t dt_1 \int_0^{t_1} dt_2 C_m(t_2)\right].
\end{equation}
Strictly, the correlation function in Eq.~(\ref{eq:gapCt}) is a quantum time
correlation function and a variety of approximate schemes exist to reconstruct
a quantum time correlation function from its classical
counterpart~\cite{Egorov1999a,Egorov1999b}.  We note that the above formalism
neglects energy transfer (or non-adiabatic effects) associated with the
intermolecular couplings $J_{mn}$.  In this $J_{mn}=0$ limit of ``pure
dephasing,'' the second order cumulant approximation is quantum mechanically
exact if the $E_m$ operators are linear in the coordinates of a harmonic
bath~\cite{Skinner1986,Reichman1996};
this linear coupling model will be adopted below.
In the more general case of energy transfer associated with population relaxation,
spectroscopy calculations require a more general approach to quantum dynamics,
which we discuss in the next section.

\section{Spectroscopy, correlation functions, and the quantum regression theorem}
\label{sec:spectroscopy}

In order to calculate spectroscopic observables, we augment our Hamiltonian
with a light-matter interaction term to be treated via time-dependent perturbation
theory~\cite{MukamelBook},
\begin{equation}
H_{\mathrm{spec}}(t)=H-\boldsymbol{\mu} \cdot \bm{E}(t);
\end{equation}
here, $\bm{E}(t)$ is a classical electric field and $\boldsymbol{\mu}$ is
the dipole operator
\begin{equation}
\boldsymbol{\mu}=\sum_{m}\mu_{m}(B^{\dagger}_{m}+B_{m})
\label{eq:dipole}
\end{equation}
where, via the Condon approximation, the dipole matrix element $\mu_m$ is
independent of the bath degrees of freedom.  For simplicity, here and
henceforth we neglect the vectorial nature of the transition dipole matrix element.

Linear absorption spectra are calculated from the
equilibrium time-correlation response function,
\begin{equation}
S^{(1)}(t) = \frac{i}{\hbar} \theta(t) \mathrm{Tr} 
	\Big\{ \left[ \mu(t), \mu \right] \rho^{\mathrm{eq}} \Big\}
        = \frac{i}{\hbar} \theta(t) \mathrm{Tr}
        \Big\{ \mu \mathcal{G}(t) \mu^\times \rho^{\mathrm{eq}} \Big\}
\end{equation}
where $\mu^\times A \equiv [\mu,A]$ and the Liouville-space propagator
is defined by $\mathcal{G}(t)A = e^{-iHt/\hbar} A e^{iHt/\hbar}$.
Two-dimensional spectra are calculated from the third-order response function,
\begin{equation}
\begin{split}
S^{(3)}(t_3,t_2,t_1) &= \left(\frac{i}{\hbar}\right)^3
    \theta(t_3)\theta(t_2)\theta(t_1) \\
    &\hspace{1em}\times \mathrm{Tr} \Big\{ \mu 
        \mathcal{G}(t_3) \mu^\times 
        \mathcal{G}(t_2) \mu^\times 
        \mathcal{G}(t_1) \mu^\times \rho^{\mathrm{eq}} \Big\}.
\end{split}
\end{equation}
With exact quantum dynamics, the above expressions produce exact spectra
that contain the effects of coherence dephasing as well as population relaxation,
with explicit treatment of system-bath correlations spanning multiple
light-matter interactions.

As a one-time observable, the linear-response function can be calculated
exactly from the time evolution of a reduced density-like operator
\begin{align}
\label{eq:rhomu}
\rho_\mu(t) &\equiv \mathrm{Tr}_\mathrm{b} \Big\{
    \mathcal{G}(t) \mu^\times \rho^{\mathrm{eq}} \Big\} 
    = \mathcal{G}_{\mathrm{red}}(t,0) \rho_\mu(0), \\
\label{eq:linear_red}
S^{(1)}(t) &= \frac{i}{\hbar} \mathrm{Tr}_{\mathrm{s}} \Big\{
    \mu \rho_\mu(t) \Big\}
    = \frac{i}{\hbar} \mathrm{Tr}_{\mathrm{s}} \Big\{
    \mu \mathcal{G}_{\mathrm{red}}(t,0) \rho_\mu(0) \Big\}.
\end{align}
Therefore, any theory of reduced dynamics may be used to calculate the linear
response function, such as a formally exact time-convolutionless quantum master
equation~\cite{Shibata1977,Chaturvedi1979,BreuerPetruccioneBook,Yoon1975,Mukamel1978},
\begin{align}
\frac{d\rho(t)}{dt} &= -\frac{i}{\hbar}\left[H_{\mathrm{s}},\rho(t)\right]
    - \mathcal{R}(t)\rho(t) \equiv -\frac{i}{\hbar} \mathcal{L}_{\mathrm{red}}(t) \rho(t), \\
\rho(t) &= T\exp\left[-\frac{i}{\hbar} 
        \int_0^t d\tau \mathcal{L}_{\mathrm{red}}(\tau) \right] \rho(0)
    \equiv \mathcal{G}_{\mathrm{red}}(t,0) \rho(0),
\end{align}
where $T$ is the time-ordering operator.
Naturally, the lowest-order approximation to the time-dependent relaxation
operator $\mathcal{R}(t)$ may be used; this is the second-order
time convolutionless (TCL2) quantum master equation~\cite{BreuerPetruccioneBook}.
Unlike the second cumulant approach described in Sec.~\ref{sec:intro}, the TCL2
master equation (detailed below) provides a perturbative description of
population relaxation and coherence dephasing on equal footing.
Importantly, in the absence of population relaxation, the TCL2
approximation yields uncoupled dephasing dynamics of the (in this case off-diagonal)
reduced operator $\sigma(t)$ that are identical to those of the second-order
cumulant approximation, and thus exact for linear coupling to a harmonic bath.
In other words, the dynamical resummation inherent in the TCL2 approximation is
exact for this example of the pure-dephasing problem (due to underlying
Gaussian statistics). We emphasize that this exactness is independent of the
number of bath degrees of freedom, the energy scales of the bath, or the
strength of the system-bath coupling.  

However, in the presence of population relaxation with $J_{mn} \neq 0$, the
TCL2 approximation is no longer exact, and its range of validity is commonly 
understood to be restricted to the nearly-Markovian, weak-coupling limit under
which it is typically derived.  We will demonstrate that such statements are
completely dependent on the observable and \textit{not} determined exclusively
by the energy scales in the Hamiltonian.  In particular, we will show that for
the same Hamiltonian, TCL2 predicts qualitatively incorrect non-equilibrium
population dynamics but quantitatively accurate linear absorption lineshapes.

Unfortunately, the simplicity of the linear response function is not 
maintained for higher-order correlation functions. 
If we were to generalize Eqs.~(\ref{eq:rhomu}) and (\ref{eq:linear_red})
to the third-order response function, we
might be led to the tempting approximation
\begin{equation}
\begin{split}
\label{eq:qrt}
\tilde{S}^{(3)}(t_3,t_2,t_1) &= \left(\frac{i}{\hbar}\right)^3
    \theta(t_3)\theta(t_2)\theta(t_1) \\
    &\hspace{-4em}\times \mathrm{Tr}_\mathrm{s} \Big\{ \mu 
        \mathcal{G}_\mathrm{red}(t_1+t_2+t_3,t_1+t_2) \\
    &\hspace{-1em} \times \mu^\times 
        \mathcal{G}_\mathrm{red}(t_1+t_2,t_1) \mu^\times 
        \mathcal{G}_\mathrm{red}(t_1,0) \rho_\mu(0) \Big\},
\end{split}
\end{equation}
which can be shown to be consistent with a quantum version of Onsager's
regression hypothesis known as the quantum regression theorem
(QRT)~\cite{Lax1963,GardinerZollerBook}.  In general, the QRT is not
exact~\cite{Ford1996,Swain1999}.  The exact multi-time correlation function is
related to the approximate one by a number of correction terms, which are
non-vanishing even to lowest-order in the system-bath
interaction~\cite{Alonso2005,Alonso2007,Goan2011}. To summarize, in the context
of the current manuscript, \textit{the rigorous calculation of nonlinear
response functions requires more information than contained in quantum master
equations}.

The violation of the QRT is intimately linked to the degree of non-Markovianity
present in the reduced dynamics.  For purely Markovian reduced dynamics, the
QRT is exact~\cite{Swain1999,Gisin1993}.  For weak system-bath coupling leading
to nearly Markovian reduced dynamics, the corrections to the QRT are small.  In
this manuscript, we will study the accuracy of TCL2 dynamics, within the
approximation implied by the QRT, Eq.~(\ref{eq:qrt}), for the simulation of
two-dimensional spectroscopy.  Unsurprisingly, we find that the results
deteriorate with increasing non-Markovianity due to stronger coupling or slower
bath degrees of freedom.

In order to maintain the simplicity of quantum master equations while seeking
broad applicability to nonlinear spectroscopy, we will evaluate the use
of the ``frozen modes'' approach, recently introduced by one of us and 
co-workers~\cite{Montoya-Castillo2015}.  The method will be described in more
detail below, but the idea is to simulate generically non-Markovian dynamics by
an average over many independent nearly-Markovian trajectories.  In each
trajectory, the low-frequency bath degrees of freedom are dynamically arrested:
they are removed from the master equation's relaxation operator and treated as
a source of static disorder.  This frozen-mode approximation is obviously best
for low-frequency (slow) degrees of freedom, which are precisely those that
contribute to non-Markovian behavior and the inaccuracy of perturbative quantum
master equations.  Compared to the original problem, each individual trajectory
in the frozen modes approach exhibits less non-Markovian behavior and weaker
system-bath coupling.  Because these are the same effects responsible for the
violation of the QRT in nonlinear response calculations, we will
demonstrate that the frozen-mode variant of TCL2 dynamics leads to accurate
two-dimensional spectra, even in highly non-Markovian regimes.

\section{Model and Methods}
\label{sec:model}

For the remainder of this work, we will adopt the common system-bath model
that assumes a harmonic nuclear bath linearly coupled to the system's excitation
number operator, i.e.~Eqs.~(\ref{eq:hamsys}) and (\ref{eq:hamsysbath}) with
\begin{align}
    H_{\mathrm{b}} &= \frac{1}{2}\sum_{m}\sum_{k}(P^{2}_{m,k}+\omega^{2}_{m,k}Q^{2}_{m,k}),
    \label{eq:hambath} \\
    E_m &= \sum_{k}c_{m,k}Q_{m,k},
    \label{eq:hamsysbathcoupling}
\end{align}
where $P_{m,k}$ and $Q_{m,k}$ are the mass-weighted momentum and position
of mode $k$ belonging to the nuclear degrees of freedom of site $m$. 
For such a simplified form of the system-bath interaction, \textit{all} 
properties are determined solely by the equilibrium autocorrelation 
function in Eq.~(\ref{eq:gapCt}),
\begin{equation}
\label{eq:Ct_harmonic}
\begin{split}
C_m(t) &= \sum_k c_{m,k}^2 \mathrm{Tr}_{\mathrm{b}} 
    \left\{ Q_{m,k}(t) Q_{m,k}(0) e^{-\beta H_{\mathrm{b}}} \right\} 
        / \mathrm{Tr}_{\mathrm{b}} e^{-\beta H_{\mathrm{b}}} \\
&\hspace{-2em} = \frac{\hbar}{\pi} \int_0^\infty d\omega J_m(\omega)
    \Big\{\coth(\beta \hbar \omega/2)\cos(\omega t) - i\sin(\omega t) \Big\},
\end{split}
\end{equation}
where we have introduced the spectral density
\begin{equation}
J_m(\omega) = \frac{\pi}{2}\sum_{k}\frac{c_{m,k}^{2}}{\omega_{m,k}}\delta(\omega-\omega_{m,k}).
\label{eq:jw}
\end{equation}
Note that the spectral density can be obtained from the
real-part of the cosine-transform of the bath correlation function
\begin{equation}
J_m(\omega) = \tanh(\beta \hbar \omega/2)
    \int_0^\infty dt \cos(\omega t)\ \mathrm{Re} C_m(t),
\end{equation}
which provides a way to extract a model spectral density from atomistic
simulations of the energy gap autocorrelation function, as done recently
for light-harvesting complexes~\cite{Valleau2012,Lee2016}.
In the following, we assume all sites have identical spectral
densities $J_m(\omega) = J(\omega)$ and employ an Ohmic spectral density with 
a high-frequency Lorentzian cutoff
\begin{equation}
J(\omega) = \frac{2\lambda\omega_{c}\omega}{\omega_{c}^{2}+\omega^{2}},
\end{equation}
which follows from the assumption of an exponentially-decaying autocorrelation
function (in the high-temperature limit), 
$\mathrm{Re} C(t) = 2\lambda k_\mathrm{B}T \exp(-\omega_c t)$.
Therefore, the bath relaxation time is given by
$\tau_\mathrm{c} = 1/\omega_\mathrm{c}$ and the magnitude of fluctuations is related to 
the reorganization energy 
$\lambda = (\hbar \pi)^{-1}\int_{0}^{\infty}d\omega J(\omega)/\omega$. 
We emphasize that most of our conclusions are not restricted to any
particular form of $J(\omega)$, including those with arbitrary structure (as
may be obtained from simulation).  Specifically, we expect that the \textit{accuracy}
of our results is only weakly dependent on the form of the spectral density, and
can be safely applied to any system-bath Hamiltonian exhibiting only linear coupling
to a harmonic bath.  However, while the method is entirely \textit{applicable} to
more general system-bath Hamiltonians, the accuracy is harder to assess.  For
example, we note that even for the pure-dephasing problem, the second cumulant
is no longer exact for systems that are quadratically-coupled to harmonic
baths~\cite{Skinner1986,Reichman1996}.

\subsection{Quantum master equations and TCL2}

Introduced above, the TCL2 quantum master equation
evolves reduced-density-like system operators $\sigma(t)$ according to the
time-local equation of motion in Eq.~(\ref{eq:tcl2}).
This equation of motion assumes a factorized initial condition of the total
density-like operator $\rho(0) = \sigma(0) \rho_\mathrm{b}(0)$, and we henceforth
assume that the initial bath density operator is at equilibrium, 
$\rho_\mathrm{b}(0) = \rho_\mathrm{b}^{\mathrm{eq}}$.
We note that for Hamiltonians with a minimum excited-state gap that is much
larger than thermal energy, the total equilibrium density operator is factorized
to an excellent approximation, 
$\rho^{\mathrm{eq}} \approx |0\rangle\langle 0| \rho_\mathrm{b}^{\mathrm{eq}}$,
justifying factorized initial conditions for equilibrium correlation functions.
In the basis of eigenstates of the system Hamiltonian, the TCL2 relaxation superoperator
$\mathcal{R}$ is a tensor with elements
\begin{equation}
\label{eq:tcl2}
\begin{split}
\mathcal{R}_{\alpha\beta\gamma\delta}(t)
    &= \mathrm{Tr}_{\mathrm{s}} \left\{ |\alpha\rangle \langle \beta|\ 
        \mathcal{R}(t) \left[ |\delta\rangle \langle \gamma | \right] \right\} \\
    &= \Gamma_{\delta\beta\alpha\gamma}^+(t) + \Gamma_{\delta\beta\alpha\gamma}^-(t) \\
        &\hspace{1em} -\delta_{\beta\delta} \sum_{\zeta} \Gamma_{\alpha\zeta\zeta\gamma}^+(t)
        -\delta_{\alpha\gamma} \sum_{\zeta} \Gamma_{\delta\zeta\zeta\beta}^-(t),
\end{split}
\end{equation}
with
\begin{equation}
\Gamma_{\alpha\beta\gamma\delta}^{\pm}(t) = \frac{1}{\hbar^2} 
    \sum_{m} N_m^{\alpha\beta} N_m^{\gamma\delta} \Theta_{m}^{\pm}(\omega_{\gamma\delta};t)
\end{equation}
and
\begin{equation}
\Theta_{m}^{\pm}(\omega;t) = \int_0^t d\tau e^{-i\omega \tau} C_m(\pm \tau).
\end{equation}
In the above, $N_m^{\alpha\beta} = \langle \alpha | B^\dagger_m B_m |
\beta\rangle$ and $C_m(t)$ is given in Eq.~(\ref{eq:Ct_harmonic}).
The relevant dimensionless parameter that controls the accuracy of
weak-coupling master equations is roughly given by $\lambda
k_{\mathrm{B}}T/(\hbar \omega_c)^2$; in particular, very slow bath relaxation
dynamics are a challenge as they violate the quasi-Markovian approximations
inherent in such master equations.  As such, we will also consider a
quantum-classical hybrid scheme that microscopically
treats slow nuclear degrees of freedom as static variables sampled from their
respective canonical distributions, discussed next.

\subsection{TCL2 master equation with frozen modes}

\begin{figure}[tbp]
\begin{center}
\includegraphics[]{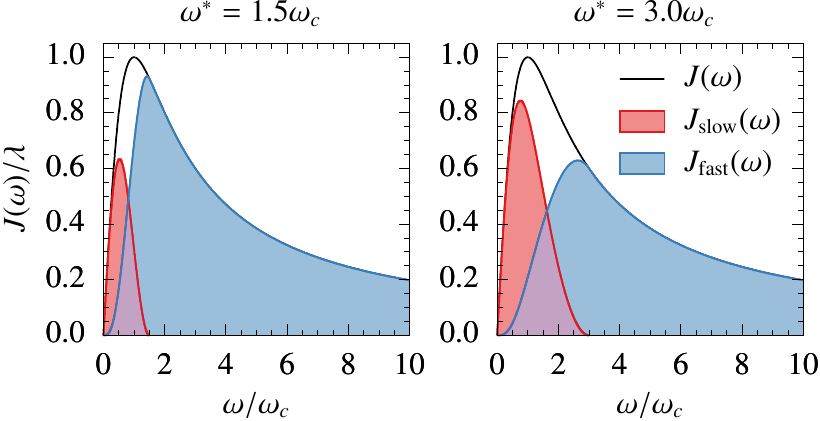}
\caption{
Two example partitionings of an Ohmic-Lorentz spectral density using different
values of the splitting frequency $\omega^{*}$; see Eqs.~(\ref{eq:jsplit}).  
Larger values of $\omega^{*}$ result in more modes being treated classically
and fewer modes treated quantum mechanically, but perturbatively.
}
\label{fig:jwsplit}
\end{center}
\end{figure}

The spectral density defined in Eq.~(\ref{eq:jw}) can formally be separated
into fast and slow components under the condition that
$J(\omega)=J_{\mathrm{fast}}(\omega)+J_{\mathrm{slow}}(\omega)$.  In the ideal
partitioning, the slow part is comprised of quasi-adiabatic modes that evolve
much more slowly than the system and can safely be treated classically, while
the fast part contains modes which evolve much more quickly than the system and
therefore can be treated with nearly-Markovian weak-coupling
theories~\cite{Thoss2001, Berkelbach2012a,Berkelbach2012b}.  The partitioning
between slow and fast parts of $J(\omega)$ can be done with a
switching function,
\begin{subequations}
\label{eq:jsplit}
\begin{align}
 J_{\mathrm{slow}}(\omega) &= s(\omega;\omega^{*}) J(\omega) \\
 J_{\mathrm{fast}}(\omega) &= [1-s(\omega;\omega^{*})] J(\omega), 
\end{align}
\end{subequations}
where $s(\omega;\omega^{*})$ is a function which goes from a value of 1 at
$\omega=0$ to a value of 0 at a specified splitting frequency $\omega=\omega^{*}$.
Following previous
work~\cite{Berkelbach2012a,Berkelbach2012b,Montoya-Castillo2015} we use the
smooth switching function
\begin{equation}
    s(\omega,\omega^{*}) = 
    \begin{cases}
        [1-(\omega/\omega^{*})^{2}]^{2}: & \omega < \omega^{*} \\ 0: & \omega \geq \omega^{*},
    \end{cases}
\end{equation}
which avoids long-time oscillatory features in the bath correlation function.  
The free parameter $\omega^{*}$ determines the set of modes to be treated classically. 
Two example partitionings of an Ohmic-Lorentz spectral density are shown in
Fig.~\ref{fig:jwsplit}.

While the slow modes could be treated with several different
semiclassical techniques, including mean-field Ehrenfest
dynamics~\cite{Thoss2001,Berkelbach2012a,Berkelbach2012b}, here we make the
simplest approximation and treat the modes described by
$J_{\mathrm{slow}}(\omega)$ as completely frozen.  In this case, positions of
the slow degrees of freedom $Q_{m,k}$ are sampled from the Boltzmann
distribution of a harmonic oscillator and used to alter the energy levels of
the system Hamiltonian,
\begin{equation}
\epsilon_m \rightarrow \epsilon_m
    + \sum_{k\in\mathrm{slow}} \bar{c}_{m,k} Q_{m,k}(0),
\end{equation}
where $\bar{c}_{m,k}$ is a renormalized coupling constant due to the partitioning
enforced by the switching function $S(\omega;\omega^{*})$.  This new system Hamiltonian
is used with the \textit{fast} part of the spectral density to perform
a single realization of TCL2 dynamics.  This process is repeated and averaged.
Further details and discussion of the method can be found in
Ref.~\onlinecite{Montoya-Castillo2015}.

In the spectroscopic context, the frozen modes are most physically interpreted
as a source of inhomogeneous broadening.  If the nuclear degrees of freedom are
extremely slow, so as to appear effectively frozen during the decay of the
dipole autocorrelation function, then this is the correct result.  We emphasize
that this inhomogeneous broadening is entirely microscopic because it
originates from degrees of freedom in the total Hamiltonian; it is not an
artificial, unidentified source of inhomogeneous broadening.  When applied
to linear absorption spectra of systems without
non-adiabatic effects, this approach is equivalent to the inhomogeneous
cumulant expansion described in Refs.~\onlinecite{Fried1993,MukamelBook},
but differs for higher-order spectroscopies.

The combination of TCL2 dynamics with frozen-mode sampling (TCL2-FM) leads to
many trajectories that are each \textit{more Markovian} than the original
problem.  This property mitigates corrections to the QRT, which are not
considered explicitly in this work.  Furthermore, the TCL2-FM approach
prevents the application of perturbation theory beyond its regime of validity
(through the use of a faster, more weakly-coupled bath in the quantum
master equation).  As
we show below, these two effects collectively produce semiquantitative
accuracy in the prediction of nonlinear spectroscopy.

\section{Results}
\label{sec:results}

\subsection{Model parameters and methodological details}

For simplicity, we present results for a system of two chromophores 
that has four accessible electronic states: the ground state, two states where
each chromophore is separately excited, and one state where both chromophores
are simultaneously excited. The absence of a quartic exciton-exciton interaction in
Eq.~(\ref{eq:hamsys}) implies that the energy of the doubly-excited state
is simply the sum of the excitation energies of the individual chromophores.
In all results, we use the electronic parameters $\epsilon_1 = 50$~cm$^{-1}$,
$\epsilon_2 = -50$~cm$^{-1}$ and $J_{12} = 100$~cm$^{-1}$, along with the
bath parameters $\omega_\mathrm{c}^{-1} = 300$~fs and $T=300$~K.
The reorganization energy $\lambda$ will be varied.
The bath frequency and temperature lead to energy scales 
$\hbar \omega_c = 18$~cm$^{-1}$ and $k_\mathrm{B}T = 208$~cm$^{-1}$,
i.e.~all energy scales are comparable, with the bath frequency being
the smallest.

Our TCL2 dynamics are generated with a fourth-order Runge-Kutta integrator.
For TCL-FM calculations, the total spectral density was discretized into 
300 modes, and those with $\omega > \omega^{*}$ were discarded.  We performed
$2\times 10^4$ realizations of frozen-mode sampling to converge
the dynamical observables.  For all frozen-mode calculations, we use a 
splitting frequency that characterizes the timescale of isolated electronic
dynamics, $\omega^* = [(\epsilon_1-\epsilon_2)^2 + 4J_{12}^2]^{1/2}/6\hbar$.
Other choices that differ by factors of order one give similar qualitative
results, and occasionally better quantitative results, but we find that the
above form is reasonable over a broad range of parameters and observables.

All approximate results will be compared to numerically exact results obtained
with the hierarchical equations of motion
(HEOM)~\cite{Tanimura1989,Ishizaki2005,Xu2007}; for the bath parameters used in
our results, the HEOM calculations required zero Matsubara frequencies
(i.e.~the high-temperature approximation with $K=0$), but as many as $L=20$
levels in the hierarchy.
For spectroscopy calculations, we use transition dipole matrix elements satisfying
$\mu_1/\mu_2 = -5$ and spectra will be presented with arbitrary units.
All calculations were performed with our open-source quantum dynamics package
\texttt{pyrho}~\cite{pyrho}.

\subsection{Linear spectroscopy and population dynamics}

\begin{figure}[tbp]
\begin{center}
\includegraphics[]{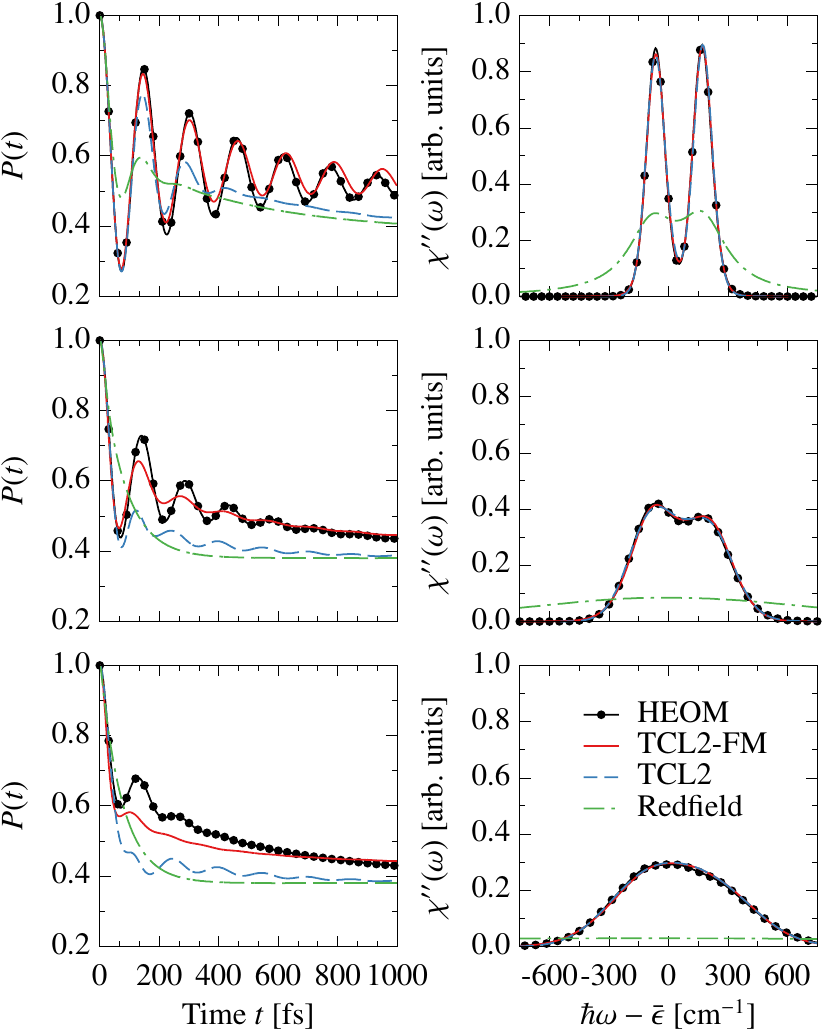}
\caption{
Population dynamics of the photo-excited higher-energy
chromophore (left column) and linear absorption spectra
(right column) with $\lambda=10$~cm$^{-1}$, 50~cm$^{-1}$, and 150~cm$^{-1}$
(top to bottom). 
}
\label{fig:abspop}
\end{center}
\end{figure}

In the frequency domain, we present the imaginary (absorptive) part of the linear-response
susceptibility,
\begin{equation}
\chi^{\prime\prime}(\omega) = \mathrm{Im} \int_0^\infty dt e^{i\omega t} S^{(1)}(t).
\end{equation}
Again we emphasize that for the model Hamiltonian adopted here, the TCL2
master equation is identical to the second-cumulant approximation and exactly
solves the pure-dephasing lineshape problem.  In the case of excited-state
electronic coupling $J\neq 0$, the TCL2 generalizes the second cumulant
approximation.  While it is not exact, the results are remarkably accurate,
as shown in the right-hand column of Fig.~\ref{fig:abspop}, even
for significant coupling to a slow bath.  
The TCL2-FM approach yields very minor quantitative improvements, which
confirms that those local-frequency modes treated as frozen are properly
interpreted as giving rise to inhomogeneous broadening.

Unlike the non-Markovian TCL2-based approaches, the Markovian Redfield
theory predicts incorrect spectra, with an accuracy that deteriorates for
increasing reorganization energy or increasing bath relaxation times.
This can be understood simply from the Markovian limit of the relevant
pure-dephasing term in the TCL2 tensor:
$1/T_2^* = \hbar^{-1} \left[ J(\omega) n(\omega) \right]_{\omega = 0}$
where $n(\omega)$ is the Bose-Einstein distribution function.  For the
Ohmic-Lorentz spectral density employed here, one finds
$1/T_2^* = 2\lambda k_{\mathrm{B}}T/\hbar \omega_c^2$.  As shown in
Fig.~\ref{fig:abspop}, this Markovian theory predicts linewidths
that are much too large.

This large discrepancy between two weak-coupling theories (Markovian
Redfield theory and non-Markovian TCL2-based theories) is quite surprising
given the perturbative nature of both approaches.  In the left-hand column
panels of Fig.~\ref{fig:abspop}, we show the population relaxation
dynamics generated by the same methods, using the initial condition
$\rho(0) = |1\rangle \langle 1| \rho_{\mathrm{b}}^{\mathrm{eq}}$.
Clearly, Redfield theory and conventional TCL2 fail in a similar manner
as the perturbation becomes large.  The TCL2-FM approach now yields results
that are quite different than vanilla TCL2, and the former predicts population
dynamics that are in good agreement with the numerically exact HEOM results.
These observations demonstrate one of our main
conclusions: the accuracy of a given dynamical technique depends on the
observable.
More specifically, TCL2-based approaches are well-suited to the evolution
of electronic coherences; we believe that this is because TCL2 reduces
to the exact second-cumulant solution for pure-dephasing problems.
Population dynamics are less accurate.  

\begin{figure*}[tbp]
\begin{center}
\includegraphics[]{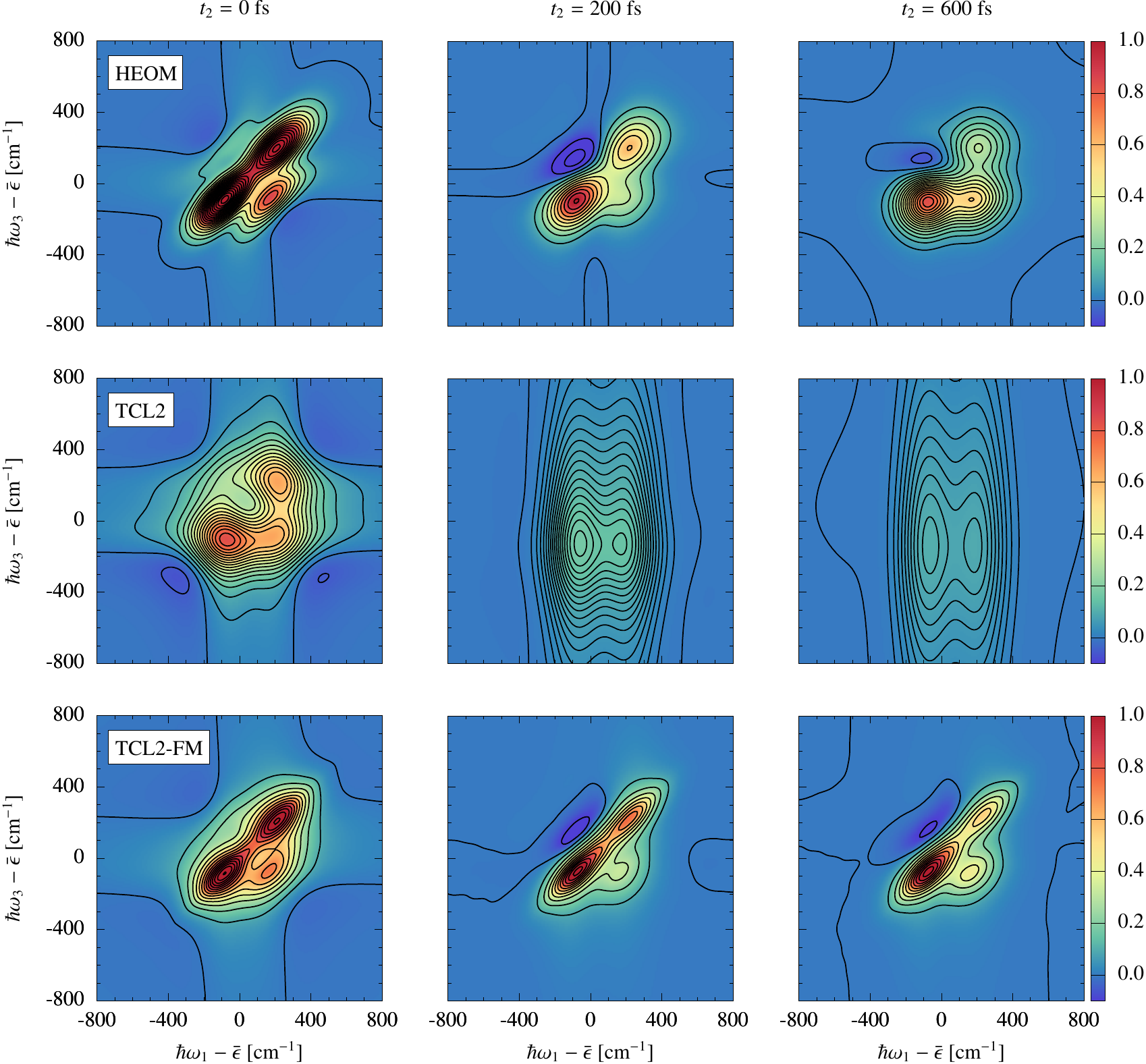}
\caption{
Two-dimensional photon echo spectra for a model dimer with
$\lambda=50$~cm$^{-1}$.  Waiting times are
shown and increase from left to right. Exact results 
(top row) are compared with the predictions of TCL2 (middle) and TCL2-FM (bottom row).  
}
\label{fig:2d}
\end{center}
\end{figure*}

\subsection{Third-order nonlinear spectroscopy}

Compared to linear response considered above, third-order nonlinear spectroscopy
is a more challenging test, due to the simultaneous importance
of coherence and population dynamics. As discussed in Sec.~\ref{sec:spectroscopy},
nonlinear spectroscopy presents an additional complication for non-Markovian quantum
master equations in the form of violations to the QRT.

We present two-dimensional electronic spectra,
obtained by Fourier transforming over the pump ($t_1$) and probe ($t_3$)
time delays, resolved by the population waiting time $t_2$.
In particular, we simulate the photon echo spectrum, generated by six
terms in the rotating wave approximation associated with rephasing (rp)
and non-rephasing (nr) pathways,
\begin{align}
A(\omega_3, t_2, \omega_1) &= \mathrm{Re} \int_0^\infty dt_1 \int_0^\infty dt_3 
    \Big\{ e^{i(\omega_1 t_1 + \omega_3 t_3)} R_{\mathrm{rp}}(t_3,t_2,t_1) \nonumber \\
    &\hspace{5em} + e^{i(-\omega_1 t_1 + \omega_3 t_3)} R_{\mathrm{nr}}(t_3,t_2,t_1) \Big\}.
\end{align}
In the above,
\begin{subequations}
\begin{align}
    R_{\textrm{rp}}(t_{3},t_{2},t_{1}) &= R_{2}(t_{3},t_{2},t_{1}) + R_{3}(t_{3},t_{2},t_{1}) + R^{*}_{1}(t_{3},t_{2},t_{1}), \\
    R_{\textrm{nr}}(t_{3},t_{2},t_{1}) &= R_{1}(t_{3},t_{2},t_{1}) + R_{4}(t_{3},t_{2},t_{1}) + R^{*}_{2}(t_{3},t_{2},t_{1}),
\end{align}
\end{subequations}
and
\begin{subequations}
\begin{align}
    R_{1}(t_{3},t_{2},t_{1}) &= \mathrm{Tr}\Big\{ \mu(t_{1}+t_{2}+t_{3})\mu(0)\rho^\mathrm{eq}\mu(t_{1})\mu(t_{1}+t_{2}) \Big\}, \\
    R_{2}(t_{3},t_{2},t_{1}) &= \mathrm{Tr}\Big\{ \mu(t_{1}+t_{2}+t_{3})\mu(0)\rho^\mathrm{eq}\mu(t_{1})\mu(t_{1}+t_{2}) \Big\}, \\
    R_{3}(t_{3},t_{2},t_{1}) &= \mathrm{Tr}\Big\{ \mu(t_{1}+t_{2}+t_{3})\mu(t_{1}+t_{2})\rho^\mathrm{eq}\mu(0)\mu(t_{1}) \Big\}, \\
    R_{4}(t_{3},t_{2},t_{1}) &= \mathrm{Tr}\Big\{ \mu(t_{1}+t_{2}+t_{3})\mu(t_{1}+t_{2})\mu(t_{1})\mu(0)\rho^\mathrm{eq} \Big\}.
\end{align}
\end{subequations}

For a moderate reorganization energy of $\lambda = 50$~cm$^{-1}$, our results
are shown in Fig.~\ref{fig:2d} for three values of the waiting time $t_2$.
Although TCL2 dynamics are qualitatively correct at $t_2=0$, this accuracy
degrades with increasing waiting times.  In particular, the spectra become
artificially broadened along the $\omega_3$ axis -- as discussed in 
Ref.~\onlinecite{Ishizaki2008} -- and completely fail to
describe the four-peak structure and regions of negativity.  In
contrast, the TCL2-FM two-dimensional spectra are in qualitative agreement with
the exact results, reproducing the appropriate peak shapes, spectral features,
and timescales.  The TCL2-FM results show an unphysical diagonal elongation
at long waiting times due to the completely frozen modes that are unable to relax.
We have found that using a smaller splitting frequency (i.e.~freezing fewer modes)
does improve the long-time results at the expense of short-time results.
This observations suggests an interesting time-dependent frozen-modes scheme
-- where modes are successively unfrozen at times longer than their characteristic
relaxation time -- though we do not pursue this approach here. 
We conclude that the computationally efficient TCL2-FM
approach yields accurate coherence and population dynamics, while mitigating
correction terms due to violation of the QRT, leading to qualitatively correct
two-dimensional spectra.  Although we only show one parameter set for simplicity,
we have tested other parameter sets and find that the TCL2-FM accuracy is
robust over a wide range of Hamiltonians.

\begin{figure}[tbp]
\begin{center}
\includegraphics[]{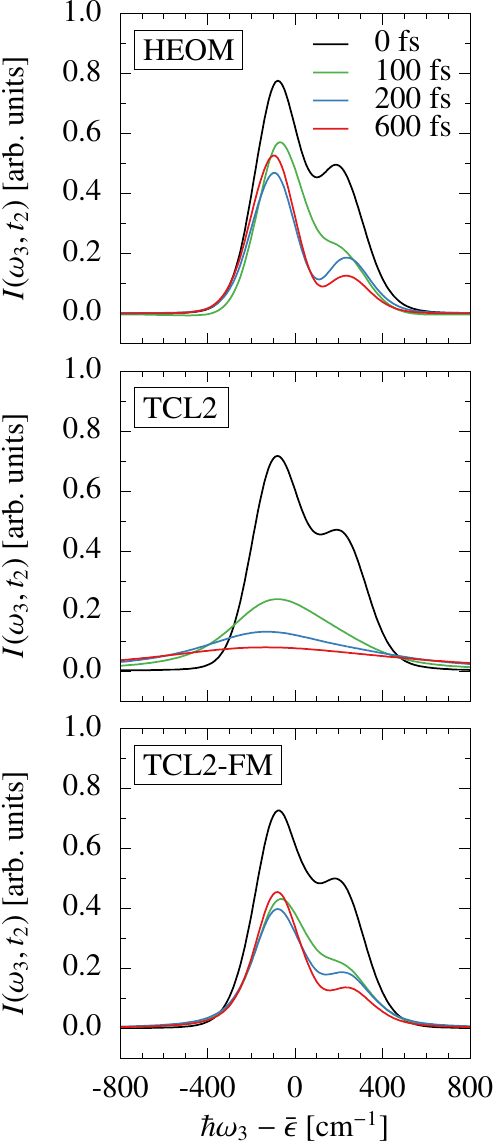}
\caption{
Pump-probe spectra $I(\omega_3,t_2)$ obtained from two-dimensional
photon echo data presented in Fig.~\ref{fig:2d}, via integration over
the $\omega_1$ axis.  Waiting times $t_2$ are given in the legend.
}
\label{fig:pumpprobe}
\end{center}
\end{figure}

For a simpler representation of the third-order response, we 
also present the time-resolved pump-probe spectra, obtained by
setting $t_1=0$ (or equivalently, integrating over $\omega_1$).
We note that the correlation functions contributing to pump-probe
spectroscopy are two-point nonequilibrium correlation functions,
which should depend sensitively on the accuracy of excited-state
population dynamics and exhibit nontrivial corrections to the 
QRT~\cite{Alonso2005,Alonso2007}. For the same parameters as in Fig.~\ref{fig:2d}, we
present the pump-probe spectra in Fig.~\ref{fig:pumpprobe}.  Consistent with
our findings for the full two-dimensional spectra, we see that TCL2 dynamics
are accurate at zero waiting time, but completely wrong for nonzero waiting
times; the spectra are much too broad and lacking multi-peaked structure.
TCL2-FM dynamics cures these deficiencies and predicts accurate pump-probe
spectra.

\section{Conclusions}

We have investigated the accuracy of the second-order time-convolutionless
(TCL2) quantum master equation for the prediction of linear and third-order
time-resolved spectroscopy.  We have argued that TCL2 dynamics generalizes the
second-cumulant approximation to problems exhibiting excited-state electronic
coupling leading to non-adiabatic dynamics and population relaxation.  For many
contemporary problems concerning the dynamics of multichromophore systems, the
bath timescales are too slow to permit classic Markovian theories of
lineshapes.  For such systems, the TCL2 approach generates nearly-exact 
linear-response spectra, even though simulations of population dynamics -- using
the same Hamiltonian parameters -- are grossly in error.

In contrast, the TCL2 approximation breaks down away from its perturbative
regime when used to simulate nonlinear spectroscopies.  We have attributed this
failure to two distinct effects:  the greater importance of population dynamics
and the violation of the quantum regression theorem (QRT).  By partitioning
microscopic bath degrees of freedom into fast and slow sets, and treating
the latter as frozen variables sampled from a thermodynamic ensemble, we have
argued that the TCL2-FM approach alleviates both of these problems.  Physically,
the TCL2-FM formalism treats modes responsible for highly non-Markovian dynamics
as a source of inhomogeneous broadening; the resulting Hamiltonian exhibits
dynamics which are consequently more Markovian and thus well-described by
weak-coupling, TCL2-type quantum master equations.
It will be interesting to implement the second-order corrections to the QRT, in
order to assess their relative importance in the prediction of nonlinear
spectra.  Work along these lines is currently in progress.

Importantly, unlike many reduced quantum dynamics techniques, the formalism of 
perturbative quantum master equations is not
tied to specific forms of the bath Hamiltonian or the system-bath interaction,
as long as multi-point equilibrium correlation functions of isolated bath operators
can be computed.
Nonetheless, the accuracy of such approaches remains to be assessed for
more generic Hamiltonians, but the absence of numerically exact results makes
this challenging.

Our findings have implications for other quantum dynamics techniques.  In
particular, we recall that the hierarchical equations of motion (HEOM) reduce
to time-convolutionless and time-convolution quantum master equations when the
hierarchy is truncated at low order~\cite{Xu2005,Xu2017}.  These low-order HEOM
calculations will exhibit the same features described here.  For linear
spectroscopy, this observation argues strongly for the use of the
time-convolutionless closure~\cite{Xu2005} of the hierarchy, in support of the
numerical observations made in Refs.~\onlinecite{Chen2009,Chen2010}.  For nonlinear
spectroscopy, low-order approximations will also violate the QRT,
and higher levels in the hierarchy will be needed in order to eliminate
these (neglected) corrections.

With regards to computational cost, the TCL2-FM approach is extremely
attractive.  In its conventional form, presented here, it scales only as
$MN_\mathrm{s}^4$, where $M$ is a parallel prefactor associated
with frozen-mode ensemble sampling and $N_\mathrm{s}$ is size of the system
Hilbert space; this scaling arises from the action of the relaxation tensor on
the reduced density matrix.  Because the frozen-modes approach leads to more
Markovian dynamics, it is tempting to explore a stochastic unraveling of the
reduced density matrix~\cite{BreuerPetruccioneBook,Zoller1987,Gisin1992,Carmichael1993,GardinerZollerBook}.
This latter approach replaces a single reduced density matrix simulation by an
average over \textit{wavefunction} dynamics with stochastic relaxation events.
This procedure would have a scaling of $L M N_{\mathrm{s}}^2$, where $L$ is a
parallel prefactor associated with the stochastic dynamics.  We
thus anticipate an accurate method which only requires many parallel
simulations of system-wavefunction dynamics, each scaling like $N_\mathrm{s}^2$
for generically non-sparse system Hamiltonians; it is hard to imagine a more
attractive computational cost with comparable qualitative accuracy over an
extremely broad range of Hamiltonian parameters.

\section*{Acknowledgments}
We thank Dr.~Roel Tempelaar and Prof.~Jim Skinner for helpful discussions.
This research was supported by start-up funds from the University of Chicago.
Calculations were performed with resources provided by the University of
Chicago Research Computing Center (RCC) and we thank Dr.~Jonathan Skone of the RCC
for assistance.


\begin{thebibliography}{58}%
\makeatletter
\providecommand \@ifxundefined [1]{%
 \@ifx{#1\undefined}
}%
\providecommand \@ifnum [1]{%
 \ifnum #1\expandafter \@firstoftwo
 \else \expandafter \@secondoftwo
 \fi
}%
\providecommand \@ifx [1]{%
 \ifx #1\expandafter \@firstoftwo
 \else \expandafter \@secondoftwo
 \fi
}%
\providecommand \natexlab [1]{#1}%
\providecommand \enquote  [1]{``#1''}%
\providecommand \bibnamefont  [1]{#1}%
\providecommand \bibfnamefont [1]{#1}%
\providecommand \citenamefont [1]{#1}%
\providecommand \href@noop [0]{\@secondoftwo}%
\providecommand \href [0]{\begingroup \@sanitize@url \@href}%
\providecommand \@href[1]{\@@startlink{#1}\@@href}%
\providecommand \@@href[1]{\endgroup#1\@@endlink}%
\providecommand \@sanitize@url [0]{\catcode `\\12\catcode `\$12\catcode
  `\&12\catcode `\#12\catcode `\^12\catcode `\_12\catcode `\%12\relax}%
\providecommand \@@startlink[1]{}%
\providecommand \@@endlink[0]{}%
\providecommand \url  [0]{\begingroup\@sanitize@url \@url }%
\providecommand \@url [1]{\endgroup\@href {#1}{\urlprefix }}%
\providecommand \urlprefix  [0]{URL }%
\providecommand \Eprint [0]{\href }%
\providecommand \doibase [0]{http://dx.doi.org/}%
\providecommand \selectlanguage [0]{\@gobble}%
\providecommand \bibinfo  [0]{\@secondoftwo}%
\providecommand \bibfield  [0]{\@secondoftwo}%
\providecommand \translation [1]{[#1]}%
\providecommand \BibitemOpen [0]{}%
\providecommand \bibitemStop [0]{}%
\providecommand \bibitemNoStop [0]{.\EOS\space}%
\providecommand \EOS [0]{\spacefactor3000\relax}%
\providecommand \BibitemShut  [1]{\csname bibitem#1\endcsname}%
\let\auto@bib@innerbib\@empty
\bibitem [{\citenamefont {Tanimura}\ and\ \citenamefont
  {Kubo}(1989)}]{Tanimura1989}%
  \BibitemOpen
  \bibfield  {author} {\bibinfo {author} {\bibfnamefont {Y.}~\bibnamefont
  {Tanimura}}\ and\ \bibinfo {author} {\bibfnamefont {R.~K.}\ \bibnamefont
  {Kubo}},\ }\href {\doibase 10.1143/JPSJ.58.101} {\bibfield  {journal}
  {\bibinfo  {journal} {J. Phys. Soc. Jpn.}\ }\textbf {\bibinfo {volume}
  {58}},\ \bibinfo {pages} {101} (\bibinfo {year} {1989})}\BibitemShut
  {NoStop}%
\bibitem [{\citenamefont {Mak}\ and\ \citenamefont {Chandler}(1990)}]{Mak1990}%
  \BibitemOpen
  \bibfield  {author} {\bibinfo {author} {\bibfnamefont {C.~H.}\ \bibnamefont
  {Mak}}\ and\ \bibinfo {author} {\bibfnamefont {D.}~\bibnamefont {Chandler}},\
  }\href {\doibase 10.1103/PhysRevA.41.5709} {\bibfield  {journal} {\bibinfo
  {journal} {Phys. Rev. A}\ }\textbf {\bibinfo {volume} {41}},\ \bibinfo
  {pages} {5709} (\bibinfo {year} {1990})}\BibitemShut {NoStop}%
\bibitem [{\citenamefont {Makarov}\ and\ \citenamefont
  {Makri}(1994)}]{Makarov1994}%
  \BibitemOpen
  \bibfield  {author} {\bibinfo {author} {\bibfnamefont {D.~E.}\ \bibnamefont
  {Makarov}}\ and\ \bibinfo {author} {\bibfnamefont {N.}~\bibnamefont
  {Makri}},\ }\href {\doibase 10.1016/0009-2614(94)00275-4} {\bibfield
  {journal} {\bibinfo  {journal} {Chem. Phys. Lett.}\ }\textbf {\bibinfo
  {volume} {221}},\ \bibinfo {pages} {482} (\bibinfo {year}
  {1994})}\BibitemShut {NoStop}%
\bibitem [{\citenamefont {Beck}\ \emph {et~al.}(2000)\citenamefont {Beck},
  \citenamefont {Jackle}, \citenamefont {Worth},\ and\ \citenamefont
  {Meyer}}]{Beck2000}%
  \BibitemOpen
  \bibfield  {author} {\bibinfo {author} {\bibfnamefont {M.}~\bibnamefont
  {Beck}}, \bibinfo {author} {\bibfnamefont {A.}~\bibnamefont {Jackle}},
  \bibinfo {author} {\bibfnamefont {G.~A.}\ \bibnamefont {Worth}}, \ and\
  \bibinfo {author} {\bibfnamefont {H.-D.}\ \bibnamefont {Meyer}},\ }\href
  {\doibase 10.1016/S0370-1573(99)00047-2} {\bibfield  {journal} {\bibinfo
  {journal} {Phys. Rep.}\ }\textbf {\bibinfo {volume} {324}},\ \bibinfo {pages}
  {1} (\bibinfo {year} {2000})}\BibitemShut {NoStop}%
\bibitem [{\citenamefont {Thoss}, \citenamefont {Wang},\ and\ \citenamefont
  {Miller}(2001)}]{Thoss2001}%
  \BibitemOpen
  \bibfield  {author} {\bibinfo {author} {\bibfnamefont {M.}~\bibnamefont
  {Thoss}}, \bibinfo {author} {\bibfnamefont {H.}~\bibnamefont {Wang}}, \ and\
  \bibinfo {author} {\bibfnamefont {W.~H.}\ \bibnamefont {Miller}},\ }\href
  {\doibase 10.1063/1.1385562} {\bibfield  {journal} {\bibinfo  {journal} {J.
  Chem. Phys.}\ }\textbf {\bibinfo {volume} {115}},\ \bibinfo {pages} {2991}
  (\bibinfo {year} {2001})}\BibitemShut {NoStop}%
\bibitem [{\citenamefont {Dunkel}, \citenamefont {Bonella},\ and\ \citenamefont
  {Coker}(2008)}]{Dunkel2008}%
  \BibitemOpen
  \bibfield  {author} {\bibinfo {author} {\bibfnamefont {E.~R.}\ \bibnamefont
  {Dunkel}}, \bibinfo {author} {\bibfnamefont {S.}~\bibnamefont {Bonella}}, \
  and\ \bibinfo {author} {\bibfnamefont {D.~F.}\ \bibnamefont {Coker}},\ }\href
  {\doibase 10.1063/1.2976441} {\bibfield  {journal} {\bibinfo  {journal} {J.
  Chem. Phys.}\ }\textbf {\bibinfo {volume} {129}},\ \bibinfo {pages} {114106}
  (\bibinfo {year} {2008})}\BibitemShut {NoStop}%
\bibitem [{\citenamefont {Chen}, \citenamefont {Cohen},\ and\ \citenamefont
  {Reichman}(2017)}]{Chen2017}%
  \BibitemOpen
  \bibfield  {author} {\bibinfo {author} {\bibfnamefont {H.-T.}\ \bibnamefont
  {Chen}}, \bibinfo {author} {\bibfnamefont {G.}~\bibnamefont {Cohen}}, \ and\
  \bibinfo {author} {\bibfnamefont {D.~R.}\ \bibnamefont {Reichman}},\ }\href
  {\doibase 10.1063/1.4974328} {\bibfield  {journal} {\bibinfo  {journal} {J.
  Chem. Phys.}\ }\textbf {\bibinfo {volume} {146}},\ \bibinfo {pages} {054105}
  (\bibinfo {year} {2017})}\BibitemShut {NoStop}%
\bibitem [{\citenamefont {Mukamel}(1995)}]{MukamelBook}%
  \BibitemOpen
  \bibfield  {author} {\bibinfo {author} {\bibfnamefont {S.}~\bibnamefont
  {Mukamel}},\ }\href@noop {} {\emph {\bibinfo {title} {{Principles of
  Nonlinear Optical Spectroscopy}}}}\ (\bibinfo  {publisher} {{Oxford
  University Press}},\ \bibinfo {year} {1995})\BibitemShut {NoStop}%
\bibitem [{\citenamefont {Hybl}\ \emph {et~al.}(1998)\citenamefont {Hybl},
  \citenamefont {Albrecht}, \citenamefont {{Gallagher Faeder}},\ and\
  \citenamefont {Jonas}}]{Hybl1998}%
  \BibitemOpen
  \bibfield  {author} {\bibinfo {author} {\bibfnamefont {J.~D.}\ \bibnamefont
  {Hybl}}, \bibinfo {author} {\bibfnamefont {A.~W.}\ \bibnamefont {Albrecht}},
  \bibinfo {author} {\bibfnamefont {S.~M.}\ \bibnamefont {{Gallagher Faeder}}},
  \ and\ \bibinfo {author} {\bibfnamefont {D.~M.}\ \bibnamefont {Jonas}},\
  }\href {\doibase 10.1016/S0009-2614(98)01140-3} {\bibfield  {journal}
  {\bibinfo  {journal} {Chem. Phys. Lett.}\ }\textbf {\bibinfo {volume}
  {297}},\ \bibinfo {pages} {307} (\bibinfo {year} {1998})}\BibitemShut
  {NoStop}%
\bibitem [{\citenamefont {Mukamel}(2000)}]{Mukamel2000}%
  \BibitemOpen
  \bibfield  {author} {\bibinfo {author} {\bibfnamefont {S.}~\bibnamefont
  {Mukamel}},\ }\href {\doibase 10.1146/annurev.physchem.51.1.691} {\bibfield
  {journal} {\bibinfo  {journal} {Annu. Rev. Phys. Chem.}\ }\textbf {\bibinfo
  {volume} {51}},\ \bibinfo {pages} {691} (\bibinfo {year} {2000})}\BibitemShut
  {NoStop}%
\bibitem [{\citenamefont {Jonas}(2003)}]{Jonas2003}%
  \BibitemOpen
  \bibfield  {author} {\bibinfo {author} {\bibfnamefont {D.~M.}\ \bibnamefont
  {Jonas}},\ }\href {\doibase 10.1146/annurev.physchem.54.011002.103907}
  {\bibfield  {journal} {\bibinfo  {journal} {Annu. Rev. Phys. Chem.}\ }\textbf
  {\bibinfo {volume} {54}},\ \bibinfo {pages} {425} (\bibinfo {year}
  {2003})}\BibitemShut {NoStop}%
\bibitem [{\citenamefont {Brixner}\ \emph {et~al.}(2005)\citenamefont
  {Brixner}, \citenamefont {Stenger}, \citenamefont {Vaswani}, \citenamefont
  {Cho}, \citenamefont {Blankenship},\ and\ \citenamefont
  {Fleming}}]{Brixner2005}%
  \BibitemOpen
  \bibfield  {author} {\bibinfo {author} {\bibfnamefont {T.}~\bibnamefont
  {Brixner}}, \bibinfo {author} {\bibfnamefont {J.}~\bibnamefont {Stenger}},
  \bibinfo {author} {\bibfnamefont {H.~M.}\ \bibnamefont {Vaswani}}, \bibinfo
  {author} {\bibfnamefont {M.}~\bibnamefont {Cho}}, \bibinfo {author}
  {\bibfnamefont {R.~E.}\ \bibnamefont {Blankenship}}, \ and\ \bibinfo {author}
  {\bibfnamefont {G.~R.}\ \bibnamefont {Fleming}},\ }\href {\doibase
  10.1038/nature03429} {\bibfield  {journal} {\bibinfo  {journal} {Nature}\
  }\textbf {\bibinfo {volume} {434}},\ \bibinfo {pages} {625} (\bibinfo {year}
  {2005})}\BibitemShut {NoStop}%
\bibitem [{\citenamefont {Engel}\ \emph {et~al.}(2007)\citenamefont {Engel},
  \citenamefont {Calhoun}, \citenamefont {Read}, \citenamefont {Ahn},
  \citenamefont {Mancal}, \citenamefont {Cheng}, \citenamefont {Blankenship},\
  and\ \citenamefont {Fleming}}]{Engel2007}%
  \BibitemOpen
  \bibfield  {author} {\bibinfo {author} {\bibfnamefont {G.~S.}\ \bibnamefont
  {Engel}}, \bibinfo {author} {\bibfnamefont {T.~R.}\ \bibnamefont {Calhoun}},
  \bibinfo {author} {\bibfnamefont {E.~L.}\ \bibnamefont {Read}}, \bibinfo
  {author} {\bibfnamefont {T.-K.}\ \bibnamefont {Ahn}}, \bibinfo {author}
  {\bibfnamefont {T.~T.}\ \bibnamefont {Mancal}}, \bibinfo {author}
  {\bibfnamefont {Y.-C.}\ \bibnamefont {Cheng}}, \bibinfo {author}
  {\bibfnamefont {R.~E.}\ \bibnamefont {Blankenship}}, \ and\ \bibinfo {author}
  {\bibfnamefont {G.~R.}\ \bibnamefont {Fleming}},\ }\href {\doibase
  10.1038/nature05678} {\bibfield  {journal} {\bibinfo  {journal} {Nature}\
  }\textbf {\bibinfo {volume} {446}},\ \bibinfo {pages} {782} (\bibinfo {year}
  {2007})}\BibitemShut {NoStop}%
\bibitem [{\citenamefont {Jang}\ \emph {et~al.}(2008)\citenamefont {Jang},
  \citenamefont {Cheng}, \citenamefont {Reichman},\ and\ \citenamefont
  {Eaves}}]{Jang2008}%
  \BibitemOpen
  \bibfield  {author} {\bibinfo {author} {\bibfnamefont {S.}~\bibnamefont
  {Jang}}, \bibinfo {author} {\bibfnamefont {Y.~C.}\ \bibnamefont {Cheng}},
  \bibinfo {author} {\bibfnamefont {D.~R.}\ \bibnamefont {Reichman}}, \ and\
  \bibinfo {author} {\bibfnamefont {J.~D.}\ \bibnamefont {Eaves}},\ }\href
  {\doibase 10.1063/1.2977974} {\bibfield  {journal} {\bibinfo  {journal} {J.
  Chem. Phys.}\ }\textbf {\bibinfo {volume} {129}},\ \bibinfo {pages} {101104}
  (\bibinfo {year} {2008})}\BibitemShut {NoStop}%
\bibitem [{\citenamefont {Plenio}\ and\ \citenamefont
  {Huelga}(2008)}]{Plenio2008}%
  \BibitemOpen
  \bibfield  {author} {\bibinfo {author} {\bibfnamefont {M.~B.}\ \bibnamefont
  {Plenio}}\ and\ \bibinfo {author} {\bibfnamefont {S.~F.}\ \bibnamefont
  {Huelga}},\ }\href {\doibase 10.1088/1367-2630/10/11/113019} {\bibfield
  {journal} {\bibinfo  {journal} {New J. Phys.}\ }\textbf {\bibinfo {volume}
  {10}},\ \bibinfo {pages} {113019} (\bibinfo {year} {2008})}\BibitemShut
  {NoStop}%
\bibitem [{\citenamefont {Mohseni}\ \emph {et~al.}(2008)\citenamefont
  {Mohseni}, \citenamefont {Rebentrost}, \citenamefont {Lloyd},\ and\
  \citenamefont {Aspuru-Guzik}}]{Mohseni2008}%
  \BibitemOpen
  \bibfield  {author} {\bibinfo {author} {\bibfnamefont {M.}~\bibnamefont
  {Mohseni}}, \bibinfo {author} {\bibfnamefont {P.}~\bibnamefont {Rebentrost}},
  \bibinfo {author} {\bibfnamefont {S.}~\bibnamefont {Lloyd}}, \ and\ \bibinfo
  {author} {\bibfnamefont {A.}~\bibnamefont {Aspuru-Guzik}},\ }\href {\doibase
  10.1063/1.3002335} {\bibfield  {journal} {\bibinfo  {journal} {J. Chem.
  Phys.}\ }\textbf {\bibinfo {volume} {129}},\ \bibinfo {pages} {174106}
  (\bibinfo {year} {2008})}\BibitemShut {NoStop}%
\bibitem [{\citenamefont {Huo}\ and\ \citenamefont {Coker}(2010)}]{Huo2010}%
  \BibitemOpen
  \bibfield  {author} {\bibinfo {author} {\bibfnamefont {P.}~\bibnamefont
  {Huo}}\ and\ \bibinfo {author} {\bibfnamefont {D.~F.}\ \bibnamefont
  {Coker}},\ }\href {\doibase 10.1063/1.3498901} {\bibfield  {journal}
  {\bibinfo  {journal} {J. Chem. Phys.}\ }\textbf {\bibinfo {volume} {133}},\
  \bibinfo {pages} {184108} (\bibinfo {year} {2010})}\BibitemShut {NoStop}%
\bibitem [{\citenamefont {Wu}\ \emph {et~al.}(2010)\citenamefont {Wu},
  \citenamefont {Liu}, \citenamefont {Shen}, \citenamefont {Cao},\ and\
  \citenamefont {Silbey}}]{Wu2010}%
  \BibitemOpen
  \bibfield  {author} {\bibinfo {author} {\bibfnamefont {J.}~\bibnamefont
  {Wu}}, \bibinfo {author} {\bibfnamefont {F.}~\bibnamefont {Liu}}, \bibinfo
  {author} {\bibfnamefont {Y.}~\bibnamefont {Shen}}, \bibinfo {author}
  {\bibfnamefont {J.}~\bibnamefont {Cao}}, \ and\ \bibinfo {author}
  {\bibfnamefont {R.~J.}\ \bibnamefont {Silbey}},\ }\href {\doibase
  10.1088/1367-2630/12/10/105012} {\bibfield  {journal} {\bibinfo  {journal}
  {New J. Phys.}\ }\textbf {\bibinfo {volume} {12}},\ \bibinfo {pages} {105012}
  (\bibinfo {year} {2010})}\BibitemShut {NoStop}%
\bibitem [{\citenamefont {Kolli}, \citenamefont {Nazir},\ and\ \citenamefont
  {Olaya-Castro}(2011)}]{Kolli2011}%
  \BibitemOpen
  \bibfield  {author} {\bibinfo {author} {\bibfnamefont {A.}~\bibnamefont
  {Kolli}}, \bibinfo {author} {\bibfnamefont {A.}~\bibnamefont {Nazir}}, \ and\
  \bibinfo {author} {\bibfnamefont {A.}~\bibnamefont {Olaya-Castro}},\ }\href
  {\doibase 10.1063/1.3652227} {\bibfield  {journal} {\bibinfo  {journal} {J.
  Chem. Phys.}\ }\textbf {\bibinfo {volume} {135}},\ \bibinfo {pages} {154112}
  (\bibinfo {year} {2011})}\BibitemShut {NoStop}%
\bibitem [{\citenamefont {Tiwari}, \citenamefont {Peters},\ and\ \citenamefont
  {Jonas}(2013)}]{Tiwari2013}%
  \BibitemOpen
  \bibfield  {author} {\bibinfo {author} {\bibfnamefont {V.}~\bibnamefont
  {Tiwari}}, \bibinfo {author} {\bibfnamefont {W.~K.}\ \bibnamefont {Peters}},
  \ and\ \bibinfo {author} {\bibfnamefont {D.~M.}\ \bibnamefont {Jonas}},\
  }\href {\doibase 10.1073/pnas.1211157110} {\bibfield  {journal} {\bibinfo
  {journal} {Proc. Natl. Acad. Sci.}\ }\textbf {\bibinfo {volume} {110}},\
  \bibinfo {pages} {1203} (\bibinfo {year} {2013})}\BibitemShut {NoStop}%
\bibitem [{\citenamefont {Tempelaar}, \citenamefont {Jansen},\ and\
  \citenamefont {Knoester}(2014)}]{Tempelaar2014}%
  \BibitemOpen
  \bibfield  {author} {\bibinfo {author} {\bibfnamefont {R.}~\bibnamefont
  {Tempelaar}}, \bibinfo {author} {\bibfnamefont {T.~L.~C.}\ \bibnamefont
  {Jansen}}, \ and\ \bibinfo {author} {\bibfnamefont {J.}~\bibnamefont
  {Knoester}},\ }\href {\doibase 10.1021/jp510074q} {\bibfield  {journal}
  {\bibinfo  {journal} {J. Phys. Chem. B}\ }\textbf {\bibinfo {volume} {118}},\
  \bibinfo {pages} {12865} (\bibinfo {year} {2014})}\BibitemShut {NoStop}%
\bibitem [{\citenamefont {Ishizaki}\ and\ \citenamefont
  {Fleming}(2009{\natexlab{a}})}]{Ishizaki2009a}%
  \BibitemOpen
  \bibfield  {author} {\bibinfo {author} {\bibfnamefont {A.}~\bibnamefont
  {Ishizaki}}\ and\ \bibinfo {author} {\bibfnamefont {G.~R.}\ \bibnamefont
  {Fleming}},\ }\href {\doibase 10.1063/1.3155214} {\bibfield  {journal}
  {\bibinfo  {journal} {J. Chem. Phys.}\ }\textbf {\bibinfo {volume} {130}},\
  \bibinfo {pages} {234110} (\bibinfo {year} {2009}{\natexlab{a}})}\BibitemShut
  {NoStop}%
\bibitem [{\citenamefont {Ishizaki}\ and\ \citenamefont
  {Fleming}(2009{\natexlab{b}})}]{Ishizaki2009b}%
  \BibitemOpen
  \bibfield  {author} {\bibinfo {author} {\bibfnamefont {A.}~\bibnamefont
  {Ishizaki}}\ and\ \bibinfo {author} {\bibfnamefont {G.~R.}\ \bibnamefont
  {Fleming}},\ }\href {\doibase 10.1063/1.3155372} {\bibfield  {journal}
  {\bibinfo  {journal} {J. Chem. Phys.}\ }\textbf {\bibinfo {volume} {130}},\
  \bibinfo {pages} {234111} (\bibinfo {year} {2009}{\natexlab{b}})}\BibitemShut
  {NoStop}%
\bibitem [{\citenamefont {Kubo}(1969)}]{Kubo1969}%
  \BibitemOpen
  \bibfield  {author} {\bibinfo {author} {\bibfnamefont {R.}~\bibnamefont
  {Kubo}},\ }\href {\doibase 10.1002/9780470143605.ch6} {\bibfield  {journal}
  {\bibinfo  {journal} {Adv. Chem. Phys.}\ }\textbf {\bibinfo {volume} {15}},\
  \bibinfo {pages} {101} (\bibinfo {year} {1969})}\BibitemShut {NoStop}%
\bibitem [{\citenamefont {Egorov}, \citenamefont {Everitt},\ and\ \citenamefont
  {Skinner}(1999)}]{Egorov1999a}%
  \BibitemOpen
  \bibfield  {author} {\bibinfo {author} {\bibfnamefont {S.~A.}\ \bibnamefont
  {Egorov}}, \bibinfo {author} {\bibfnamefont {K.~F.}\ \bibnamefont {Everitt}},
  \ and\ \bibinfo {author} {\bibfnamefont {J.~L.}\ \bibnamefont {Skinner}},\
  }\href {\doibase 10.1021/jp9919314} {\bibfield  {journal} {\bibinfo
  {journal} {J. Phys. Chem. A}\ }\textbf {\bibinfo {volume} {103}},\ \bibinfo
  {pages} {9494} (\bibinfo {year} {1999})}\BibitemShut {NoStop}%
\bibitem [{\citenamefont {Egorov}, \citenamefont {Rabani},\ and\ \citenamefont
  {Berne}(1999)}]{Egorov1999b}%
  \BibitemOpen
  \bibfield  {author} {\bibinfo {author} {\bibfnamefont {S.~A.}\ \bibnamefont
  {Egorov}}, \bibinfo {author} {\bibfnamefont {E.}~\bibnamefont {Rabani}}, \
  and\ \bibinfo {author} {\bibfnamefont {B.~J.}\ \bibnamefont {Berne}},\ }\href
  {\doibase 10.1021/jp9921349} {\bibfield  {journal} {\bibinfo  {journal} {J.
  Phys. Chem. B}\ }\textbf {\bibinfo {volume} {103}},\ \bibinfo {pages} {10978}
  (\bibinfo {year} {1999})}\BibitemShut {NoStop}%
\bibitem [{\citenamefont {Skinner}\ and\ \citenamefont
  {Hsu}(1986)}]{Skinner1986}%
  \BibitemOpen
  \bibfield  {author} {\bibinfo {author} {\bibfnamefont {J.~L.}\ \bibnamefont
  {Skinner}}\ and\ \bibinfo {author} {\bibfnamefont {D.}~\bibnamefont {Hsu}},\
  }\href {\doibase 10.1021/j100412a013} {\bibfield  {journal} {\bibinfo
  {journal} {J. Phys. Chem.}\ }\textbf {\bibinfo {volume} {90}},\ \bibinfo
  {pages} {4931} (\bibinfo {year} {1986})}\BibitemShut {NoStop}%
\bibitem [{\citenamefont {Reichman}, \citenamefont {Silbey},\ and\
  \citenamefont {Su{\'{a}}rez}(1996)}]{Reichman1996}%
  \BibitemOpen
  \bibfield  {author} {\bibinfo {author} {\bibfnamefont {D.}~\bibnamefont
  {Reichman}}, \bibinfo {author} {\bibfnamefont {R.~J.}\ \bibnamefont
  {Silbey}}, \ and\ \bibinfo {author} {\bibfnamefont {A.}~\bibnamefont
  {Su{\'{a}}rez}},\ }\href {\doibase 10.1063/1.472976} {\bibfield  {journal}
  {\bibinfo  {journal} {J. Chem. Phys.}\ }\textbf {\bibinfo {volume} {105}},\
  \bibinfo {pages} {10500} (\bibinfo {year} {1996})}\BibitemShut {NoStop}%
\bibitem [{\citenamefont {Shibata}, \citenamefont {Takahashi},\ and\
  \citenamefont {Hashitsume}(1977)}]{Shibata1977}%
  \BibitemOpen
  \bibfield  {author} {\bibinfo {author} {\bibfnamefont {F.}~\bibnamefont
  {Shibata}}, \bibinfo {author} {\bibfnamefont {Y.}~\bibnamefont {Takahashi}},
  \ and\ \bibinfo {author} {\bibfnamefont {N.}~\bibnamefont {Hashitsume}},\
  }\href {\doibase 10.1007/BF01040100} {\bibfield  {journal} {\bibinfo
  {journal} {J. Stat. Phys.}\ }\textbf {\bibinfo {volume} {17}},\ \bibinfo
  {pages} {171} (\bibinfo {year} {1977})}\BibitemShut {NoStop}%
\bibitem [{\citenamefont {Chaturvedi}\ and\ \citenamefont
  {Shibata}(1979)}]{Chaturvedi1979}%
  \BibitemOpen
  \bibfield  {author} {\bibinfo {author} {\bibfnamefont {S.}~\bibnamefont
  {Chaturvedi}}\ and\ \bibinfo {author} {\bibfnamefont {F.}~\bibnamefont
  {Shibata}},\ }\href {\doibase 10.1007/BF01319852} {\bibfield  {journal}
  {\bibinfo  {journal} {Z. Phys. B Condens. Matter}\ }\textbf {\bibinfo
  {volume} {35}},\ \bibinfo {pages} {297} (\bibinfo {year} {1979})}\BibitemShut
  {NoStop}%
\bibitem [{\citenamefont {Breuer}\ and\ \citenamefont
  {Petruccione}(2007)}]{BreuerPetruccioneBook}%
  \BibitemOpen
  \bibfield  {author} {\bibinfo {author} {\bibfnamefont {H.-P.}\ \bibnamefont
  {Breuer}}\ and\ \bibinfo {author} {\bibfnamefont {F.}~\bibnamefont
  {Petruccione}},\ }\href@noop {} {\emph {\bibinfo {title} {{The Theory of Open
  Quantum Systems}}}}\ (\bibinfo  {publisher} {{Oxford University Press}},\
  \bibinfo {year} {2007})\BibitemShut {NoStop}%
\bibitem [{\citenamefont {Yoon}, \citenamefont {Deutch},\ and\ \citenamefont
  {Freed}(1975)}]{Yoon1975}%
  \BibitemOpen
  \bibfield  {author} {\bibinfo {author} {\bibfnamefont {B.}~\bibnamefont
  {Yoon}}, \bibinfo {author} {\bibfnamefont {J.~M.}\ \bibnamefont {Deutch}}, \
  and\ \bibinfo {author} {\bibfnamefont {J.~H.}\ \bibnamefont {Freed}},\ }\href
  {\doibase 10.1063/1.430417} {\bibfield  {journal} {\bibinfo  {journal} {J.
  Chem. Phys.}\ }\textbf {\bibinfo {volume} {62}},\ \bibinfo {pages} {4687}
  (\bibinfo {year} {1975})}\BibitemShut {NoStop}%
\bibitem [{\citenamefont {Mukamel}, \citenamefont {Oppenheim},\ and\
  \citenamefont {Ross}(1978)}]{Mukamel1978}%
  \BibitemOpen
  \bibfield  {author} {\bibinfo {author} {\bibfnamefont {S.}~\bibnamefont
  {Mukamel}}, \bibinfo {author} {\bibfnamefont {I.}~\bibnamefont {Oppenheim}},
  \ and\ \bibinfo {author} {\bibfnamefont {J.}~\bibnamefont {Ross}},\ }\href
  {\doibase 10.1103/PhysRevA.17.1988} {\bibfield  {journal} {\bibinfo
  {journal} {Phys. Rev. A}\ }\textbf {\bibinfo {volume} {17}},\ \bibinfo
  {pages} {1988} (\bibinfo {year} {1978})}\BibitemShut {NoStop}%
\bibitem [{\citenamefont {Lax}(1963)}]{Lax1963}%
  \BibitemOpen
  \bibfield  {author} {\bibinfo {author} {\bibfnamefont {M.}~\bibnamefont
  {Lax}},\ }\href {\doibase 10.1103/PhysRev.129.2342} {\bibfield  {journal}
  {\bibinfo  {journal} {Phys. Rev.}\ }\textbf {\bibinfo {volume} {129}},\
  \bibinfo {pages} {2342} (\bibinfo {year} {1963})}\BibitemShut {NoStop}%
\bibitem [{\citenamefont {Gardiner}\ and\ \citenamefont
  {Zoller}(2004)}]{GardinerZollerBook}%
  \BibitemOpen
  \bibfield  {author} {\bibinfo {author} {\bibfnamefont {P.}~\bibnamefont
  {Gardiner}}\ and\ \bibinfo {author} {\bibfnamefont {P.}~\bibnamefont
  {Zoller}},\ }\href@noop {} {\emph {\bibinfo {title} {{Quantum Noise}}}}\
  (\bibinfo  {publisher} {{Springer-Verlag, Berlin}},\ \bibinfo {year}
  {2004})\BibitemShut {NoStop}%
\bibitem [{\citenamefont {Ford}\ and\ \citenamefont
  {O'Connell}(1996)}]{Ford1996}%
  \BibitemOpen
  \bibfield  {author} {\bibinfo {author} {\bibfnamefont {G.~W.}\ \bibnamefont
  {Ford}}\ and\ \bibinfo {author} {\bibfnamefont {R.}~\bibnamefont
  {O'Connell}},\ }\href {\doibase 10.1103/PhysRevLett.77.798} {\bibfield
  {journal} {\bibinfo  {journal} {Phys. Rev. Lett.}\ }\textbf {\bibinfo
  {volume} {77}},\ \bibinfo {pages} {798} (\bibinfo {year} {1996})}\BibitemShut
  {NoStop}%
\bibitem [{\citenamefont {Swain}(1999)}]{Swain1999}%
  \BibitemOpen
  \bibfield  {author} {\bibinfo {author} {\bibfnamefont {S.}~\bibnamefont
  {Swain}},\ }\href {\doibase 10.1088/0305-4470/14/10/013} {\bibfield
  {journal} {\bibinfo  {journal} {J. Phys. A. Math. Gen.}\ }\textbf {\bibinfo
  {volume} {14}},\ \bibinfo {pages} {2577} (\bibinfo {year}
  {1999})}\BibitemShut {NoStop}%
\bibitem [{\citenamefont {Alonso}\ and\ \citenamefont
  {de~Vega}(2005)}]{Alonso2005}%
  \BibitemOpen
  \bibfield  {author} {\bibinfo {author} {\bibfnamefont {D.}~\bibnamefont
  {Alonso}}\ and\ \bibinfo {author} {\bibfnamefont {I.}~\bibnamefont
  {de~Vega}},\ }\href {\doibase 10.1103/PhysRevLett.94.200403} {\bibfield
  {journal} {\bibinfo  {journal} {Phys. Rev. Lett.}\ }\textbf {\bibinfo
  {volume} {94}},\ \bibinfo {pages} {1} (\bibinfo {year} {2005})}\BibitemShut
  {NoStop}%
\bibitem [{\citenamefont {Alonso}\ and\ \citenamefont
  {de~Vega}(2007)}]{Alonso2007}%
  \BibitemOpen
  \bibfield  {author} {\bibinfo {author} {\bibfnamefont {D.}~\bibnamefont
  {Alonso}}\ and\ \bibinfo {author} {\bibfnamefont {I.}~\bibnamefont
  {de~Vega}},\ }\href {\doibase 10.1103/PhysRevA.75.052108} {\bibfield
  {journal} {\bibinfo  {journal} {Phys. Rev. A - At. Mol. Opt. Phys.}\ }\textbf
  {\bibinfo {volume} {75}},\ \bibinfo {pages} {052108} (\bibinfo {year}
  {2007})}\BibitemShut {NoStop}%
\bibitem [{\citenamefont {Goan}, \citenamefont {Chen},\ and\ \citenamefont
  {Jian}(2011)}]{Goan2011}%
  \BibitemOpen
  \bibfield  {author} {\bibinfo {author} {\bibfnamefont {H.~S.}\ \bibnamefont
  {Goan}}, \bibinfo {author} {\bibfnamefont {P.~W.}\ \bibnamefont {Chen}}, \
  and\ \bibinfo {author} {\bibfnamefont {C.~C.}\ \bibnamefont {Jian}},\ }\href
  {\doibase 10.1063/1.3570581} {\bibfield  {journal} {\bibinfo  {journal} {J.
  Chem. Phys.}\ }\textbf {\bibinfo {volume} {134}} (\bibinfo {year} {2011}),\
  10.1063/1.3570581}\BibitemShut {NoStop}%
\bibitem [{\citenamefont {Gisin}(1993)}]{Gisin1993}%
  \BibitemOpen
  \bibfield  {author} {\bibinfo {author} {\bibfnamefont {N.}~\bibnamefont
  {Gisin}},\ }\href {\doibase 10.1080/09500349314552331} {\bibfield  {journal}
  {\bibinfo  {journal} {J. Mod. Opt.}\ }\textbf {\bibinfo {volume} {40}},\
  \bibinfo {pages} {2313} (\bibinfo {year} {1993})}\BibitemShut {NoStop}%
\bibitem [{\citenamefont {Montoya-Castillo}, \citenamefont {Berkelbach},\ and\
  \citenamefont {Reichman}(2015)}]{Montoya-Castillo2015}%
  \BibitemOpen
  \bibfield  {author} {\bibinfo {author} {\bibfnamefont {A.}~\bibnamefont
  {Montoya-Castillo}}, \bibinfo {author} {\bibfnamefont {T.~C.}\ \bibnamefont
  {Berkelbach}}, \ and\ \bibinfo {author} {\bibfnamefont {D.~R.}\ \bibnamefont
  {Reichman}},\ }\href {\doibase 10.1063/1.4935443} {\bibfield  {journal}
  {\bibinfo  {journal} {J. Chem. Phys.}\ }\textbf {\bibinfo {volume} {143}},\
  \bibinfo {pages} {194198} (\bibinfo {year} {2015})}\BibitemShut {NoStop}%
\bibitem [{\citenamefont {Valleau}, \citenamefont {Eisfeld},\ and\
  \citenamefont {Aspuru-Guzik}(2012)}]{Valleau2012}%
  \BibitemOpen
  \bibfield  {author} {\bibinfo {author} {\bibfnamefont {S.}~\bibnamefont
  {Valleau}}, \bibinfo {author} {\bibfnamefont {A.}~\bibnamefont {Eisfeld}}, \
  and\ \bibinfo {author} {\bibfnamefont {A.}~\bibnamefont {Aspuru-Guzik}},\
  }\href {\doibase 10.1063/1.4769079} {\bibfield  {journal} {\bibinfo
  {journal} {J. Chem. Phys.}\ }\textbf {\bibinfo {volume} {137}},\ \bibinfo
  {pages} {224103} (\bibinfo {year} {2012})}\BibitemShut {NoStop}%
\bibitem [{\citenamefont {Lee}\ and\ \citenamefont {Coker}(2016)}]{Lee2016}%
  \BibitemOpen
  \bibfield  {author} {\bibinfo {author} {\bibfnamefont {M.~K.}\ \bibnamefont
  {Lee}}\ and\ \bibinfo {author} {\bibfnamefont {D.~F.}\ \bibnamefont
  {Coker}},\ }\href {\doibase 10.1021/acs.jpclett.6b01440} {\bibfield
  {journal} {\bibinfo  {journal} {J. Phys. Chem. Lett.}\ }\textbf {\bibinfo
  {volume} {7}},\ \bibinfo {pages} {3171} (\bibinfo {year} {2016})}\BibitemShut
  {NoStop}%
\bibitem [{\citenamefont {Berkelbach}, \citenamefont {Reichman},\ and\
  \citenamefont {Markland}(2012)}]{Berkelbach2012a}%
  \BibitemOpen
  \bibfield  {author} {\bibinfo {author} {\bibfnamefont {T.~C.}\ \bibnamefont
  {Berkelbach}}, \bibinfo {author} {\bibfnamefont {D.~R.}\ \bibnamefont
  {Reichman}}, \ and\ \bibinfo {author} {\bibfnamefont {T.~E.}\ \bibnamefont
  {Markland}},\ }\href {\doibase 10.1063/1.3671372} {\bibfield  {journal}
  {\bibinfo  {journal} {J. Chem. Phys.}\ }\textbf {\bibinfo {volume} {136}},\
  \bibinfo {pages} {034113} (\bibinfo {year} {2012})}\BibitemShut {NoStop}%
\bibitem [{\citenamefont {Berkelbach}, \citenamefont {Markland},\ and\
  \citenamefont {Reichman}(2012)}]{Berkelbach2012b}%
  \BibitemOpen
  \bibfield  {author} {\bibinfo {author} {\bibfnamefont {T.~C.}\ \bibnamefont
  {Berkelbach}}, \bibinfo {author} {\bibfnamefont {T.~E.}\ \bibnamefont
  {Markland}}, \ and\ \bibinfo {author} {\bibfnamefont {D.~R.}\ \bibnamefont
  {Reichman}},\ }\href {\doibase 10.1063/1.3687342} {\bibfield  {journal}
  {\bibinfo  {journal} {J. Chem. Phys.}\ }\textbf {\bibinfo {volume} {136}},\
  \bibinfo {pages} {084104} (\bibinfo {year} {2012})}\BibitemShut {NoStop}%
\bibitem [{\citenamefont {Fried}\ and\ \citenamefont
  {Mukamel}(1993)}]{Fried1993}%
  \BibitemOpen
  \bibfield  {author} {\bibinfo {author} {\bibfnamefont {L.~E.}\ \bibnamefont
  {Fried}}\ and\ \bibinfo {author} {\bibfnamefont {S.}~\bibnamefont
  {Mukamel}},\ }\href {\doibase 10.1002/9780470143605.ch6} {\bibfield
  {journal} {\bibinfo  {journal} {Adv. Chem. Phys.}\ }\textbf {\bibinfo
  {volume} {84}},\ \bibinfo {pages} {435} (\bibinfo {year} {1993})}\BibitemShut
  {NoStop}%
\bibitem [{\citenamefont {Ishizaki}\ and\ \citenamefont
  {Tanimura}(2005)}]{Ishizaki2005}%
  \BibitemOpen
  \bibfield  {author} {\bibinfo {author} {\bibfnamefont {A.}~\bibnamefont
  {Ishizaki}}\ and\ \bibinfo {author} {\bibfnamefont {Y.}~\bibnamefont
  {Tanimura}},\ }\href {\doibase 10.1143/JPSJ.74.3131} {\bibfield  {journal}
  {\bibinfo  {journal} {J. Phys. Soc. Japan}\ }\textbf {\bibinfo {volume}
  {74}},\ \bibinfo {pages} {3131} (\bibinfo {year} {2005})}\BibitemShut
  {NoStop}%
\bibitem [{\citenamefont {Xu}\ and\ \citenamefont {Yan}(2007)}]{Xu2007}%
  \BibitemOpen
  \bibfield  {author} {\bibinfo {author} {\bibfnamefont {R.-X.}\ \bibnamefont
  {Xu}}\ and\ \bibinfo {author} {\bibfnamefont {Y.}~\bibnamefont {Yan}},\
  }\href {\doibase 10.1103/PhysRevE.75.031107} {\bibfield  {journal} {\bibinfo
  {journal} {Phys. Rev. E}\ }\textbf {\bibinfo {volume} {75}},\ \bibinfo
  {pages} {031107} (\bibinfo {year} {2007})}\BibitemShut {NoStop}%
\bibitem [{pyr()}]{pyrho}%
  \BibitemOpen
  \href@noop {} {\enquote {\bibinfo {title} {pyrho: a python package for
  reduced density matrix techniques},}\ }\bibinfo {howpublished}
  {\url{https://github.com/berkelbach-group/pyrho}}\BibitemShut {NoStop}%
\bibitem [{\citenamefont {Ishizaki}\ and\ \citenamefont
  {Tanimura}(2008)}]{Ishizaki2008}%
  \BibitemOpen
  \bibfield  {author} {\bibinfo {author} {\bibfnamefont {A.}~\bibnamefont
  {Ishizaki}}\ and\ \bibinfo {author} {\bibfnamefont {Y.}~\bibnamefont
  {Tanimura}},\ }\href {\doibase 10.1016/j.chemphys.2007.10.037} {\bibfield
  {journal} {\bibinfo  {journal} {Chem. Phys.}\ }\textbf {\bibinfo {volume}
  {347}},\ \bibinfo {pages} {185} (\bibinfo {year} {2008})}\BibitemShut
  {NoStop}%
\bibitem [{\citenamefont {Xu}\ \emph {et~al.}(2005)\citenamefont {Xu},
  \citenamefont {Cui}, \citenamefont {Li}, \citenamefont {Mo},\ and\
  \citenamefont {Yan}}]{Xu2005}%
  \BibitemOpen
  \bibfield  {author} {\bibinfo {author} {\bibfnamefont {R.~X.}\ \bibnamefont
  {Xu}}, \bibinfo {author} {\bibfnamefont {P.}~\bibnamefont {Cui}}, \bibinfo
  {author} {\bibfnamefont {X.~Q.}\ \bibnamefont {Li}}, \bibinfo {author}
  {\bibfnamefont {Y.}~\bibnamefont {Mo}}, \ and\ \bibinfo {author}
  {\bibfnamefont {Y.}~\bibnamefont {Yan}},\ }\href {\doibase 10.1063/1.1850899}
  {\bibfield  {journal} {\bibinfo  {journal} {J. Chem. Phys.}\ }\textbf
  {\bibinfo {volume} {122}},\ \bibinfo {pages} {041103} (\bibinfo {year}
  {2005})}\BibitemShut {NoStop}%
\bibitem [{\citenamefont {Xu}\ \emph {et~al.}(2017)\citenamefont {Xu},
  \citenamefont {Song}, \citenamefont {Song},\ and\ \citenamefont
  {Shi}}]{Xu2017}%
  \BibitemOpen
  \bibfield  {author} {\bibinfo {author} {\bibfnamefont {M.}~\bibnamefont
  {Xu}}, \bibinfo {author} {\bibfnamefont {L.}~\bibnamefont {Song}}, \bibinfo
  {author} {\bibfnamefont {K.}~\bibnamefont {Song}}, \ and\ \bibinfo {author}
  {\bibfnamefont {Q.}~\bibnamefont {Shi}},\ }\href {\doibase 10.1063/1.4974926}
  {\bibfield  {journal} {\bibinfo  {journal} {J. Chem. Phys.}\ }\textbf
  {\bibinfo {volume} {146}},\ \bibinfo {pages} {064102} (\bibinfo {year}
  {2017})}\BibitemShut {NoStop}%
\bibitem [{\citenamefont {Chen}\ \emph {et~al.}(2009)\citenamefont {Chen},
  \citenamefont {Zheng}, \citenamefont {Shi},\ and\ \citenamefont
  {Yan}}]{Chen2009}%
  \BibitemOpen
  \bibfield  {author} {\bibinfo {author} {\bibfnamefont {L.}~\bibnamefont
  {Chen}}, \bibinfo {author} {\bibfnamefont {R.}~\bibnamefont {Zheng}},
  \bibinfo {author} {\bibfnamefont {Q.}~\bibnamefont {Shi}}, \ and\ \bibinfo
  {author} {\bibfnamefont {Y.}~\bibnamefont {Yan}},\ }\href {\doibase
  10.1063/1.3213013} {\bibfield  {journal} {\bibinfo  {journal} {J. Chem.
  Phys.}\ }\textbf {\bibinfo {volume} {131}},\ \bibinfo {pages} {094502}
  (\bibinfo {year} {2009})}\BibitemShut {NoStop}%
\bibitem [{\citenamefont {Chen}\ \emph {et~al.}(2010)\citenamefont {Chen},
  \citenamefont {Zheng}, \citenamefont {Shi},\ and\ \citenamefont
  {Yan}}]{Chen2010}%
  \BibitemOpen
  \bibfield  {author} {\bibinfo {author} {\bibfnamefont {L.}~\bibnamefont
  {Chen}}, \bibinfo {author} {\bibfnamefont {R.}~\bibnamefont {Zheng}},
  \bibinfo {author} {\bibfnamefont {Q.}~\bibnamefont {Shi}}, \ and\ \bibinfo
  {author} {\bibfnamefont {Y.}~\bibnamefont {Yan}},\ }\href {\doibase
  10.1063/1.3293039} {\bibfield  {journal} {\bibinfo  {journal} {J. Chem.
  Phys.}\ }\textbf {\bibinfo {volume} {132}},\ \bibinfo {pages} {024505}
  (\bibinfo {year} {2010})}\BibitemShut {NoStop}%
\bibitem [{\citenamefont {Zoller}, \citenamefont {Marte},\ and\ \citenamefont
  {Walls}(1987)}]{Zoller1987}%
  \BibitemOpen
  \bibfield  {author} {\bibinfo {author} {\bibfnamefont {P.}~\bibnamefont
  {Zoller}}, \bibinfo {author} {\bibfnamefont {M.}~\bibnamefont {Marte}}, \
  and\ \bibinfo {author} {\bibfnamefont {D.}~\bibnamefont {Walls}},\ }\href
  {\doibase 10.1103/PhysRevA.35.198} {\bibfield  {journal} {\bibinfo  {journal}
  {Phys. Rev. A}\ }\textbf {\bibinfo {volume} {35}},\ \bibinfo {pages} {198}
  (\bibinfo {year} {1987})}\BibitemShut {NoStop}%
\bibitem [{\citenamefont {Gisin}\ and\ \citenamefont
  {Percival}(1992)}]{Gisin1992}%
  \BibitemOpen
  \bibfield  {author} {\bibinfo {author} {\bibfnamefont {N.}~\bibnamefont
  {Gisin}}\ and\ \bibinfo {author} {\bibfnamefont {I.~C.}\ \bibnamefont
  {Percival}},\ }\href {\doibase 10.1088/0305-4470/25/21/023} {\bibfield
  {journal} {\bibinfo  {journal} {J. Phys. A. Math. Gen.}\ }\textbf {\bibinfo
  {volume} {25}},\ \bibinfo {pages} {5677} (\bibinfo {year}
  {1992})}\BibitemShut {NoStop}%
\bibitem [{\citenamefont {Carmichael}(1993)}]{Carmichael1993}%
  \BibitemOpen
  \bibfield  {author} {\bibinfo {author} {\bibfnamefont {H.~J.}\ \bibnamefont
  {Carmichael}},\ }\href {\doibase 10.1103/PhysRevLett.70.2273} {\bibfield
  {journal} {\bibinfo  {journal} {Phys. Rev. Lett.}\ }\textbf {\bibinfo
  {volume} {70}},\ \bibinfo {pages} {2273} (\bibinfo {year}
  {1993})}\BibitemShut {NoStop}%
\end{thebibliography}
\end{document}